\documentclass[conference,a4paper]{APSIPA2023}
\usepackage{multirow}
\usepackage{graphicx}
\usepackage{amsmath}
\usepackage{amssymb}
\usepackage{amsxtra}
\usepackage{threeparttable}
\usepackage{cite}

\renewcommand{\baselinestretch}{0.94}

\usepackage{geometry}
\usepackage{fancyhdr}

\fancypagestyle{firststyle} {
    \fancyhf{}
    \fancyhead[C]{2023 Asia Pacific Signal and Information Processing Association Annual Summit and Conference (APSIPA ASC)}

}

\begin{document}

\title{Simultaneous Measurement {of} Multiple Acoustic Attributes Using Structured Periodic Test Signals Including Music {and} Other Sound Materials}

\author{%
\authorblockN{%
Hideki Kawahara\authorrefmark{1}, 
Kohei Yatabe\authorrefmark{2}, 
Ken-Ichi Sakakibara\authorrefmark{3}, 
Mitsunori Mizumachi\authorrefmark{4}, 
Tatsuya Kitamura\authorrefmark{5}, 
}
\authorblockA{%
\authorrefmark{1}
Wakayama University, Japan  \ \
E-mail: kawahara@wakayama-u.ac.jp~\hspace{-13mm}}
\authorblockA{%
\authorrefmark{2}
Tokyo University of Agriculture and Technology, Japan \ \
E-mail: yatabe@go.tuat.ac.jp~\hspace{50.5mm}\hspace{-12mm}}
\authorblockA{%
\authorrefmark{3}
Health Sciences University of Hokkaido, Japan \ \
E-mail: kis@hoku-iryo-u.ac.jp~\hspace{36.5mm}\hspace{-12mm}}
\authorblockA{%
\authorrefmark{4}
Kyushu Institute of Technology, Japan \ \
E-mail: mizumach@ecs.kyutech.ac.jp~\hspace{13.5mm}\hspace{-12mm}}
\authorblockA{%
\authorrefmark{5}
Konan University, Japan \ \
E-mail: t-kitamu@konan-u.ac.jp~\hspace{1.5mm}\hspace{-12mm}}
}

\maketitle
\thispagestyle{firststyle}
\pagestyle{fancy}

\begin{abstract}
We introduce a general framework for measuring acoustic properties such as liner time-invariant (LTI) response, signal-dependent time-invariant (SDTI) component, and random and time-varying (RTV) component simultaneously using structured periodic test signals.
The framework also enables music pieces and other sound materials as test signals by ``safeguarding'' them by adding slight deterministic ``noise.''
Measurement using swept-sin, MLS (Maxim Length Sequence), and their variants are special cases of the proposed framework.
We implemented interactive and real-time measuring tools based on this framework and made them open-source.
Furthermore, we applied this framework to assess pitch extractors objectively. 
\end{abstract}

\section{Introduction}
Computing devices today are $10^9$ times more powerful than those available a half-century ago~\cite{Leiserson2020science}.
This power makes it possible to process huge amount of speech materials (for example:~\cite{Kahn2020icassp,Mehrish2023informationFusion}).
This advancement motivated us to establish a framework for making speech materials more usable based on our recent findings~\cite{Kawahara2020apsipa,Kawahara2021icassp,Kawahara2022ast}.

The speech materials mentioned above do not necessarily fulfill recommended conditions for scientific research~\cite{Patel2018speechHearing}.
However, the abovementioned conditions are too strict depending on the research purpose.
There is room to make speech materials usable by providing tools for assisting data acquisition and retrospective assesment~\cite{sakakibara2020jasj}.

The framework we introduce in this article provides a solid basis for tools making speech materials reusable.
We start with a big picture of the framework, followed by descriptions of constituent algorithms.
Then, we introduce acoustic measurement tools followed by applications other than acoustic measurement.
Finally, we discuss issues and related works.

The main contribution of this paper is the fundamental reformulation of the underlying algorithms and the complete revision of the infrastructure, computationally efficient implementation based on FFT and inverse FFT.
Our previous works consisted of many conceptual confusions and ad hoc and inefficient procedures~\cite{Kawahara2019apsipa,Kawahara2020apsipa,Kawahara2021icassp,Kawahara2021isfb,Kawahara2021isST,Kawahara2021apsipafb,Kawahara2022ast,Kawahara2022isFoMes}.
This paper enables us to renew all applications and tools while maintaining their functionality.

\section{Measurement of acoustic attributes}
\begin{figure}[tbp]
\begin{center}
\includegraphics[width=\hsize]{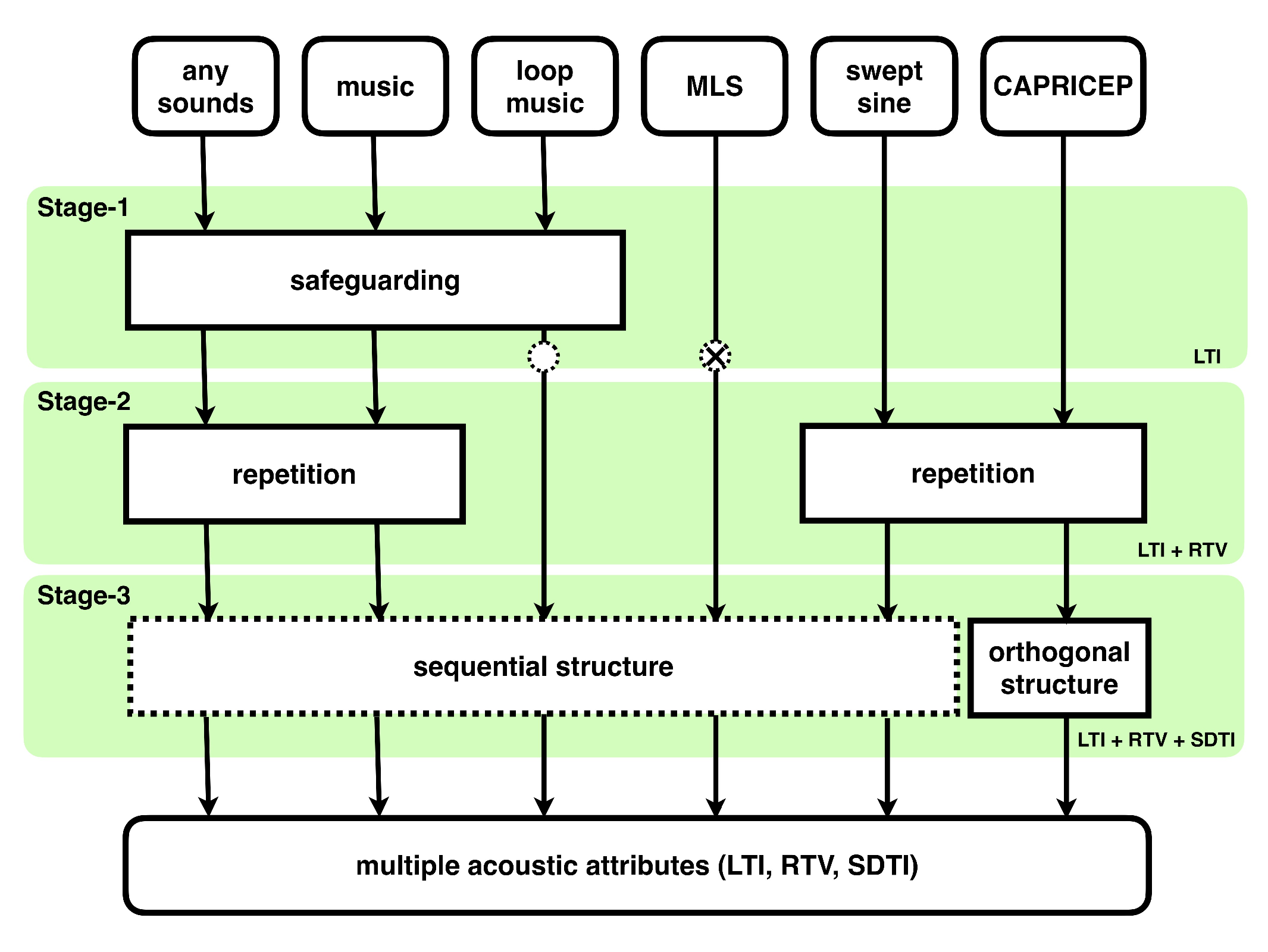}\\ 
\vspace{-3mm}
\caption{Acoustic attributes and test signals. (LTI: Linear Time-Invariant, RTV: Random and Time-Varying, and SDTI: Signal Dependent Time-Invariant). The terminal symbol placed at the bottom of Stage-1 on the line from ``MLS'' represents that results are erroneous. The terminal symbol on the line from ``loop music'' represents that it destroys the signal design, looping.}
\label{fig:measurementFrmwk}
\end{center}
\end{figure}
Figure~\ref{fig:measurementFrmwk} shows a schematic representation of the whole types of acoustic attributes measurement that we will introduce in this article.
The following three subsections outline this framework followed by sections describing technical details and applications.

\subsection{Linear time-invariant (LTI) response: Stage-1}
For linear time invariant (LTI) systems, response to input impulse (impulse response) uniquely determines the target system.
In the frequency domain, dividing the Fourier transform of the target system output by the Fourier transform of the input signal provides the transfer function of the target system.
The symbol ``swept sine''~\cite{Aoshima1981jasa} and ``MLS'' (Maximum Length Sequence~\cite{Borish1983aesj,Rife1989aesj}) are commonly used test signals.
They are members of TSP (Time Stretched Pulse) and have flat power spectrum.
The symbol ``CAPRICEP'' (Casceded All-Pass filters with RandomIzed CEnter frequencies and Phase polority) is also a new member of TSP we proposed~\cite{Kawahara2021icassp}.

Spectral division using any other signals also provides the transfer function, in principle.
However, their spectrum generally consists of very weak component(s) that makes spectral division impractical.
We introduced ``safeguarding'' method for making such signal usable for transfer function measurement~\cite{Kawahara2022ast}.

\subsection{Random and time-varying component: Stage-2}
Two neighboring periods excerpted from the repeatedly concatenated periodic segments are identical.
However, periods excerpted from different parts of the acquired signal are not identical because the background noise and sensitivity fluctuations do not have the same periodicity.
The difference between observed periods provides background noise and sensitivity fluctuations information.
This way, the repetitive presentation of periodic segments makes simultaneous measurement of the LTI response and the disturbance in the measurement.

\subsection{Signal dependent component: Stage-3}
Estimated impulse responses using different input signals are identical when the system is strictly LTI.
However, estimated impulse responses of an acoustic system in the real world are not identical for different input signals because of non-linearity, even after suppressing the random and time-varying component by averaging many repetitive measurement results.
The differences in estimated impulse responses are the second type of disturbing component.
The component depends on the used test signals. It is necessary to describe the used test signals.

Repetition (used in Stage-2) is one type of structuring for the test signal.
Sequentially aligning different repetitive test signals is the second type of test signal structuring.
Using CAPRICEP, we introduce the third type of structuring, simultaneous structuring, by using the orthogonal nature of the Walsh-Hadamard matrix.

\section{Algorithm}
This section describes the underlying algorithms in each stage.
The test and the acquired signals are discrete-time signals and share the same sampling clock.
(In the appendix, we introduce a solution for handling signals sampled by independent clocks.)

\subsection{LTI response: Stage-1, linear convolution}
\begin{figure}[tbp]
\begin{center}
\includegraphics[width=\hsize]{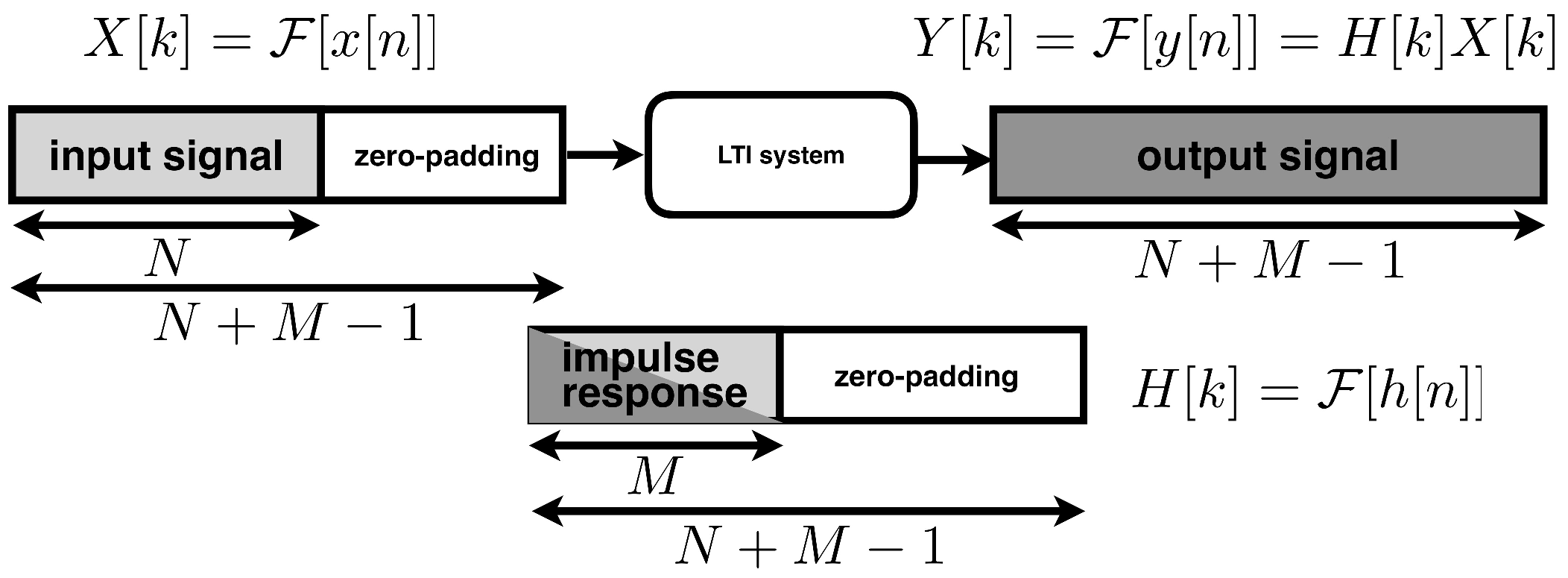}
\caption{Convolution of discrete-time signals. Implementation using cyclic convolution with zero-padding.}
\label{fig:ltiLinear}
\end{center}
\end{figure}

Let $x[n], y[n],$ and $h[n]$ represent an input signal, an output signal, and the impulse response of a system.
The time index $n$ is an integer.
Assume that $x[n]$ is non-zero for $0\le n \le N-1$ and $h[n]$ is non-zero for $0\le n \le M-1$.
Then, the output signal $y[n]$ is non-zero for $0\le n \le M+N-1$.
This setting, usually called as ``zero-padding'', is a common practice to calculate linear convolution using DFT-based circular convolution~\cite{Oppenheim1989prentice}.
(See Fig.~\ref{fig:ltiLinear})

By using the discrete Fourier transform $\mathcal{F}[\cdot ]$ of length $L>M+N-1$, the following equation provides the transfer function $H[k]$ of the system.
The symbol $k$ represents the index of the discrete frequency.
\begin{align}\label{eq:transfLin}
H[k] & = \frac{\mathcal{F}\left[y[n]\right]}{\mathcal{F}\left[x[n]\right]} .
\end{align}

Without additive noise, the inverse transform $\mathcal{F}^{-1}[\cdot ]$ of the transfer function provides the estimate of the impulse response $h_\mathrm{est}[n]$.
\begin{align}\label{eq:transfLin}
h_\mathrm{est}[n] & = \mathcal{F}^{-1}\left[H[k]\right] .
\end{align}

The typical TSP (Time Stretched-Pulse) signal, swept-sine, and the recent member of TSP, unit-CAPRICEP, (usually) have frequency independent gain, such as $\left|H[k]\right|=1$ for all $k$.
Because of the background noise, frequency-dependent gain improves estimation accuracy~\cite{Nakahara2021Jaes}.

\subsubsection{Terminal symbols}
Figure~\ref{fig:measurementFrmwk} has two terminal symbols for ``loop music'' and ``MLS'' at the bottom of Stage-1.
We placed the symbol because MLS is a periodic sequence and is orthogonal only under cyclic convolution.
The separated single cycle of the MLS sequence is not a time-stretched pulse under linear convolution.
For loop music, isolating one repeated phrase is usable for calculating the LTI response, although it is not looping and destroys the original design purpose.

\subsubsection{Signal safeguarding}
Equation~(\ref{eq:transfLin}) holds for input signals with frequency-dependent spectral values.
However, the estimated transfer functions are not usable because small absolute values in the denominator magnify the effects of observation noise.
Signal-safeguarding, our proposal, significantly suppresses observation noise effects and makes any sounds appropriate for acoustic measurement~\cite{Kawahara2022ast}.
With safeguarding, if necessary, all signals are relevant for acoustic measurement.

\subsection{LTI response, and random and time-varying component: Stage-2, cyclic convolution}
\begin{figure}[tbp]
\begin{center}
\includegraphics[width=\hsize]{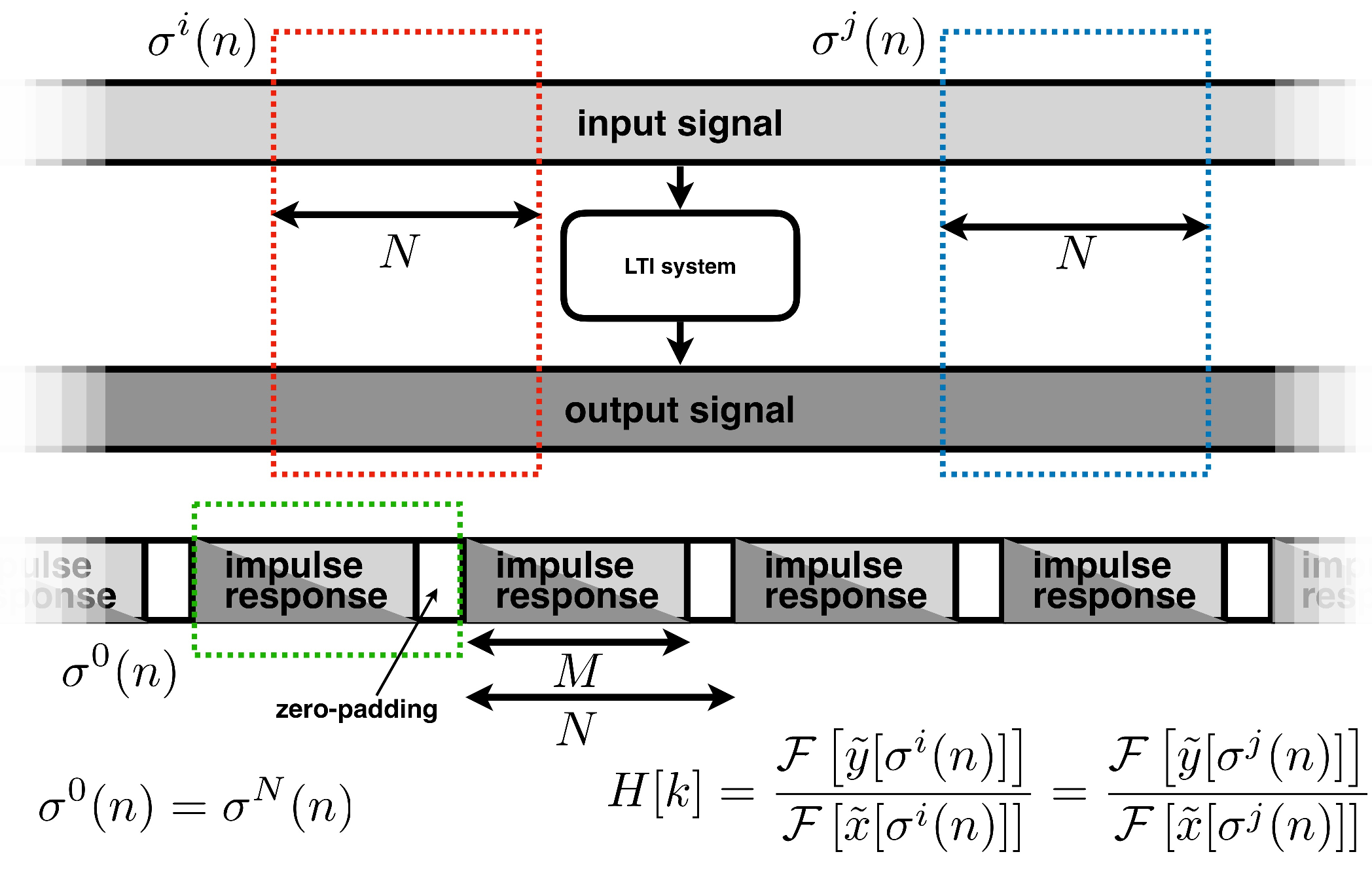}\\ 
\vspace{-3mm}
\caption{Convolution of periodic signals. Implementation using cyclic convolution without zero-padding for input and  output signals. The same cyclic permutation to input and output does not change the estimated transfer function. }
\label{fig:ltiPeriodic}
\end{center}
\end{figure}
Figure~\ref{fig:ltiPeriodic} summarizes this stage.
Zero padding is inefficient.
The periodic nature of the discrete Fourier transform removes the zero-padding for input and output in Fig.~\ref{fig:ltiLinear}.

Assume that $x[n], y[n],$ and $h[n]$ are periodic and share the same period, $L$.
Then, there is no need for excessive zero-padding for test signals.
Using a swept-sine signal in this repetitive presentation, removing zero-valued parts does not cause any problems.
The following equation provides the transfer function.
\begin{align}
H[k] & = \frac{\mathcal{F}\left[\tilde{y}[n]\right]}{\mathcal{F}\left[\tilde{x}[n]\right]} ,
\end{align}
where $\tilde{x}[n]$ and $\tilde{y}[n]$ represent that the signals are periodic.

Periodic test signal enables repeated measurement.
Disturbance in measurement, such as background noise and interfering sounds, does not share the same periodicity with the test signal.
Then, it is safe to assume that disturbing sounds are independent at each repetition.
Averaging of the estimated impulse responses reduces the estimation variance.
This averaging and calculation of observed standard deviation provide an estimate of the random and time-varying component (RTV in Fig.~\ref{fig:measurementFrmwk}).

Periodic test signal has additional merits.
Because it is periodic, there is no need to repeat the calculation of the Fourier transform of the test signal.
Because it is periodic, there is no need to find the initial position of the analysis segment.
Wherever the initial position is, the signal is periodic as far as the length is identical to the period.
In short, the following holds.
\begin{align}
\frac{\mathcal{F}\left[\tilde{y}[\sigma^i(n)]\right]}{\mathcal{F}\left[\tilde{x}[\sigma^i(n)]\right]}
 & = \frac{\mathcal{F}\left[\tilde{y}[\sigma^j(n)]\right]}{\mathcal{F}\left[\tilde{x}[\sigma^j(n)]\right]} ,
\end{align}
where $\sigma^i(n)$ and $\sigma^j(n)$ represent $i$-th and $j$-th cyclic permutations of the discrete time sequence $n$.
Figure~\ref{fig:ltiPeriodic} illustrates these merits.

\subsection{LTI response, and random and time-varying, and signal dependent components: Stage-3, using different test signals}
\begin{figure}[tbp]
\begin{center}
\includegraphics[width=\hsize]{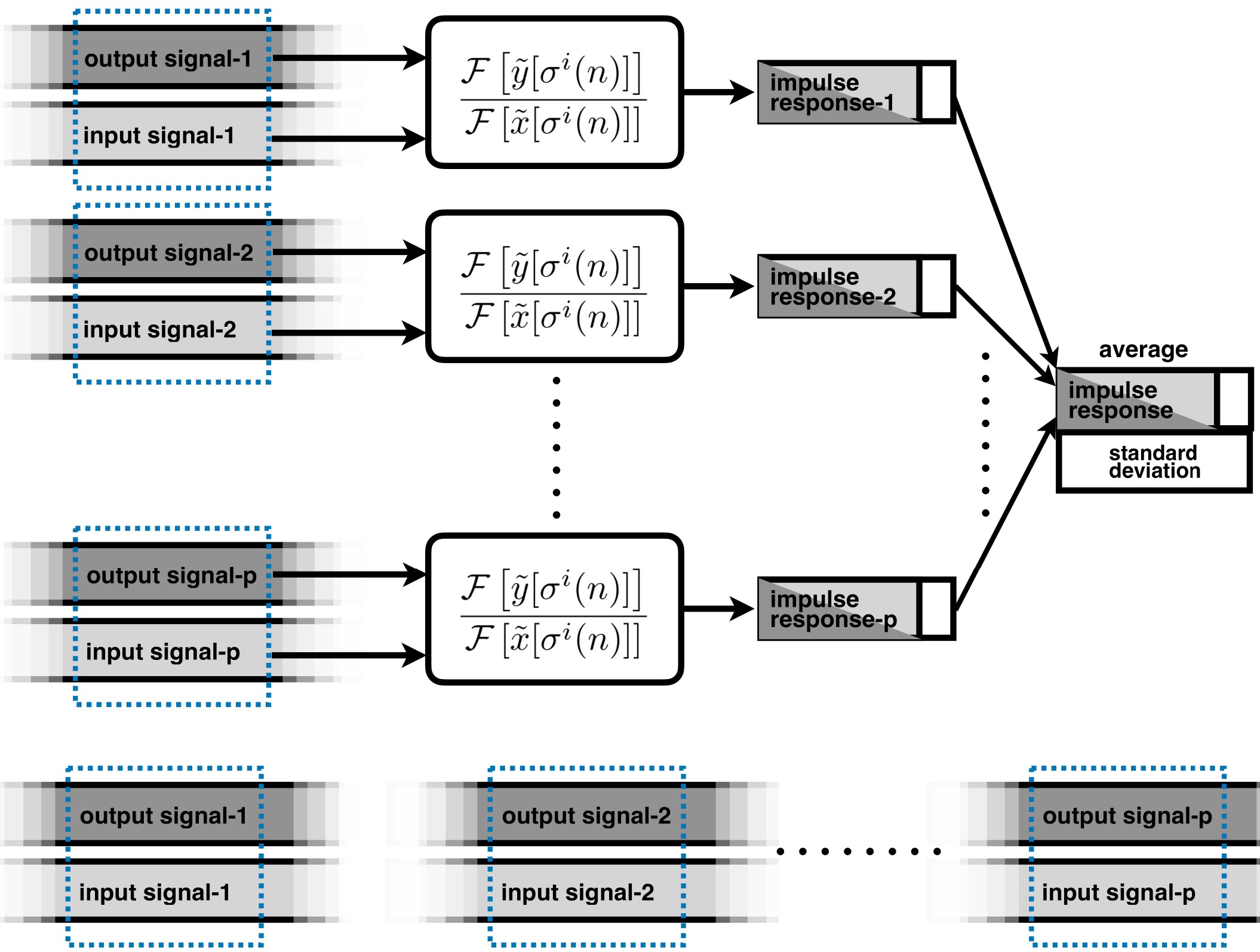}
\caption{Serial structuring illustration. Signal-dependent component isolation by averaging and calculating sample standard deviation of estimated impulse responses using different test signals. }
\label{fig:sigDepSerial}
\end{center}
\end{figure}
The estimated impulse response is identical when the system is LTI strictly, and no noise exists.
However, in actual acoustic measurement, impulse responses estimated using different input test signals are different.
Averaging impulse responses using many different test signals provides less distorted LTI response.
The difference between each impulse response and the averaged one provides an estimate of the magnitude of signal-dependent deviation.
Note that we assume each estimated impulse response for a test signal is an average of repetition, and this averaging suppresses random variations.

Figure~\ref{fig:sigDepSerial} illustrates one implementation of this strategy, serial structuring.
This serial structuring is the only option for test signals other than CAPRICEP.

\subsection{LTI response, and random and time-varying, and signal dependent components: Stage-3, structured test signal}
CAPRICEP has another option, orthogonal structuring.
Each unit-CAPRICEP has a raised-cosine-shaped power envelope.
Periodic allocation of each unit-CAPRICEP with 50\% overlap yields a constant power envelope.
We use the Walsh-Hadamard matrix $B$ of order 4 to determine the polarity of three different unit-CAPRICEPs.
\begin{align}
B & = 
\left[\!\!
\begin{array}{rrrr}
     1  &   1  &   1  &   1 \\
     1  &  -1  &  1  & -1 \\
     1  &   1  &  -1  &  -1 \\
     1 &   -1  &  -1  &   1 
\end{array}\!\!
\right]\! = \!\left[\!\!
\begin{array}{llll}
    b_{1,1}   &  \cdots  &  \cdots  &  b_{1,4} \\
     \vdots &  b_{2,2}  &     & \vdots \\
     \vdots &      &  \ddots  &  \vdots \\
     b_{4,1} &   \cdots  &  \cdots  &   b_{4,4} 
\end{array}
\!\!\!\right] .
\end{align}

\subsubsection{Periodic signal generation}
The following part describes a procedure to make a periodic test signal $\tilde{s}[n]$ from three different unit-CAPRICEPs.
The discrete Fourier transform of $\tilde{s}[n]$ yields discrete spectral representation $S[k]$.
The following procedure makes the absolute value $|S[k]|$ a constant, that is independent of the discrete frequency $k$.

Multiplying coefficient $b_{i,j}$ to each unit-CAPRICEP $g_\mathrm{uC}^{(q)}[n]$ and applying the overlap-and-add operation with $N/4$ shift for three different unit-CAPRICEPs provide a base unit sequence $u[n]$.
\begin{align}
u[n] = \sum_{q=1}^3 C_q \sum_{r=1}^4 b_{q,r}\  g_\mathrm{uC}^{(q)}\!\!\left[n-\frac{(q-1)N}{4}\right]  
\end{align}
where the super-script $q$ of $g_\mathrm{uC}^{(q)}[n]$ is the identifier of each unit-CAPRICEP.
The coefficient $C_q$ value is 1 for $q=1,2,$ and $\sqrt{2}$ for $q=3$.

The next step is to make a periodic test signal $\tilde{s}[n]$ from the base unit sequence $u[n]$.
Since the width of a unit-CAPRICEP is four to six times wider than the allocation interval $N/4$ (see Appendix), it is necessary to allocate (overlap-and-add) $u[n]$ more than $6+1$ times to make a periodic segment.
Let $P$ represent the number of repetitions.
Considering the condition and for simplicity, we select an even number $P \ge 8$.

The periodic test signal $\tilde{s}[n]$ is an excerpt from the following intermediate signal $s_\mathrm{tmp}[n]$.
\begin{align}
s_\mathrm{tmp}[n] = \sum_{p=1}^{P} u[n-(p-1)N] ,
\end{align}
Excerpting a segment with length $N$ around the center location of $s_\mathrm{tmp}[n]$ provides the periodic test signal $\tilde{s}[n]$.
\begin{align}
\tilde{s}[n] & = s_\mathrm{tmp}\!\!\left[n-\left(\frac{P}{2}-1\right)N\right] ,\ \ (n = 0, \ldots, N\!-\!1)
\end{align}

\subsubsection{Simultaneous multiple impulse response measurement}
The following procedure provides three different impulse responses from the output signal $\tilde{y}[n]$.
First, similar to the second stage, using the ratio of the discrete Fourier transform of input test signal and output provides an estimate of the impulse response $\hat{h}_\mathrm{L}[n]$.
\begin{align}
\hat{h}_\mathrm{L}[n] & = \mathcal{F}^{-1}\!\!\left[\frac{\mathcal{F}[\tilde{y}[n]]}{\mathcal{F}[\tilde{s}[n]]}\right] ,
\end{align}
where the subscript $L$ of $\hat{h}_\mathrm{L}[n]$ represents that it is a longer impulse response obtained from the measurement.
The length of $\hat{h}_\mathrm{L}[n]$ is $L$.
Although, in actual measurement, other than the initial part, the noise floor (due to background noise and non-linearity, mainly) masks the impulse response corresponds to the LTI part.

The orthogonal structure of $\tilde{s}[n]$ provides three shorter impulse responses by using the virtual target signal $\tilde{v}_\mathrm{S}^{(q)}[n]$.
\begin{align}
\tilde{v}_\mathrm{S}^{(q)}[n] & = \sum_{r=1}^4 b_{q,r}\  \delta\left[n, \frac{(q-1)N}{4}\right] ,
\end{align}
where $\delta[i,j]$ is the Kronecker delta and $q$ identifies the corresponding unit-CAPRICEP.

The first $N/4$ elements of the signal defined by the following equation provide a short impulse response $\hat{h}_\mathrm{S}^{(q)}[n]$.
\begin{align}
\hat{h}_\mathrm{S}^{(q)}[n] & = \mathcal{F}^{-1}\!\!\left[\frac{\mathcal{F}[\tilde{v}_\mathrm{S}^{(q)}[n]]\mathcal{F}[\tilde{y}[n]]}{\mathcal{F}[\tilde{s}[n]]}\right] .
\end{align}

As noted, the initial $N/4$ elements are unique and relevant for impulse response estimates.
Deviations from the same initial part of $\hat{h}_\mathrm{L}[n]$ represent the signal-dependent component.

\begin{figure}[tbp]
\begin{center}
\includegraphics[width=\hsize]{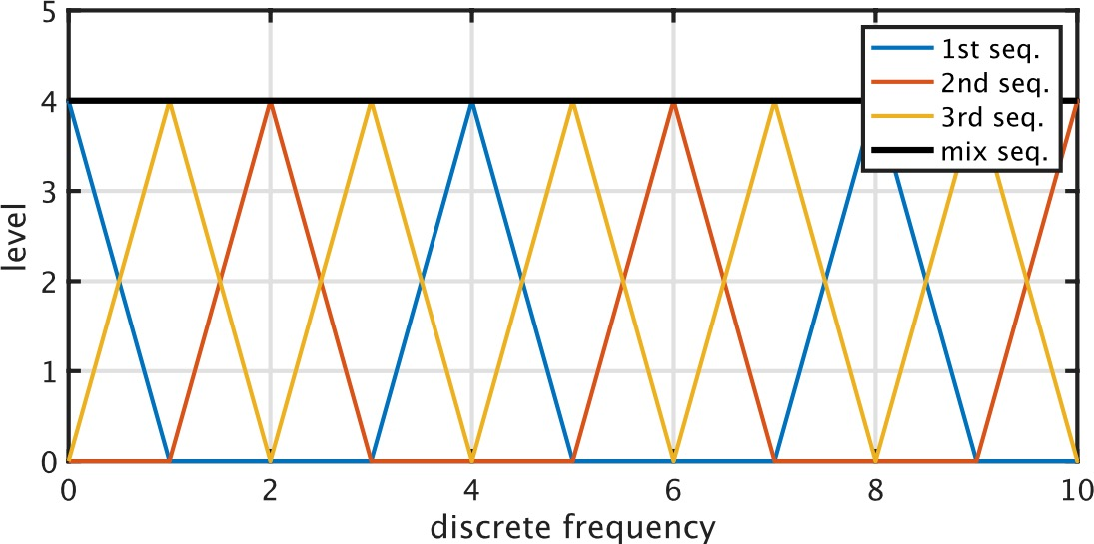}\\ 
\vspace{-2mm}
\caption{Discrete spectral levels of the whole test signal $\tilde{s}[n]$ and constituent sequences corresponding first, second, and the third unit-CAPRICEP. 
The plot shows the initial 11 discrete frequency components.}
\label{fig:mixCapSeq4}
\end{center}
\end{figure}

Note that $\mathcal{F}[\tilde{v}_\mathrm{S}^{(q)}[n]]$ functions as a selector of discrete frequency components.
Figure~\ref{fig:mixCapSeq4} illustrates how it works in the discrete frequency domain.
For example, $\hat{h}_\mathrm{S}^{(1)}[n]$ selects the bleu-colored component, and $\hat{h}_\mathrm{S}^{(2)}[n]$ and $\hat{h}_\mathrm{S}^{(3)}[n]$ select red-colored and yellow-colored components.

There is a caveat that the procedure above suppresses the signal-dependent component resulting from even-symmetric nonlinearity.
It is necessary to estimate impulse responses using the negated test signal $-\tilde{s}[n]$ to detect components caused by even-symmetric nonlinearity.

\subsection{Implementation details}
The descriptions mentioned above are simplified summaries.
We placed descriptions of technical details in Appendix sections for readability.
The discussed details are as follows:
a) Optimized weighting shape for reducing artifacts due to truncation of acquired interfering signals. (Appendix~\ref{ss:truncation})
b) Phase manipulation function that makes better localization of the generated unit-CAPRICEP. (Appndix~\ref{ss:phaseMan})
c) Evaluation of signal safeguarding merits. (Appendix~\ref{ss:safeguarding})

\section{Acoustic measurement tools}
We developed acoustic measurement tools based on the proposed framework.
This section introduces several examples.
The tools introduced here are accessible in the first author's GitHub repositories~\cite{kawahara2020gitHk}.

\subsection{Control panel for acoustic applications}
\begin{figure}[tbp]
\begin{center}
\includegraphics[width=\hsize]{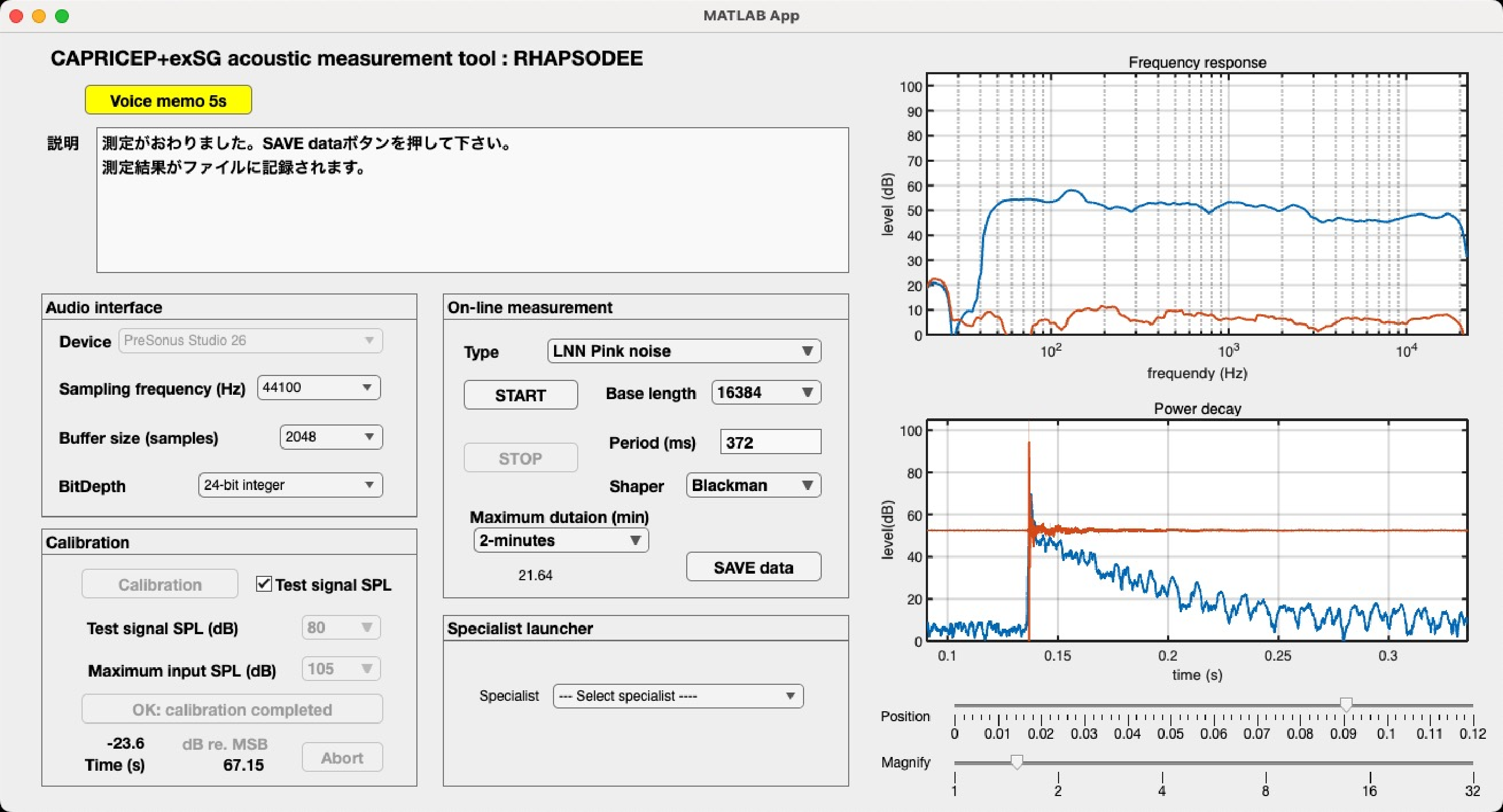}
\caption{Snapshot of the GUI of the ``Control panel'' for acoustic calibration and application launcher~\cite{Kawahara2023smac}.}
\label{fig:acousticCalibTool}
\end{center}
\end{figure}
Figure~\ref{fig:acousticCalibTool} shows a snapshot of the GUI of the ``Control panel'' for acoustic calibration and application launcher~\cite{Kawahara2023smac}.
The top-left wide field is for showing prompt instructions for users.
The left-middle and bottom sub-panels are for setting the audio interface and calibrating input sensitivity.
The center sub-panel is for controlling the acoustic measurement.
The center-bottom sub-panel is the application launcher.

\begin{figure}[tbp]
\begin{center}
\includegraphics[width=\hsize]{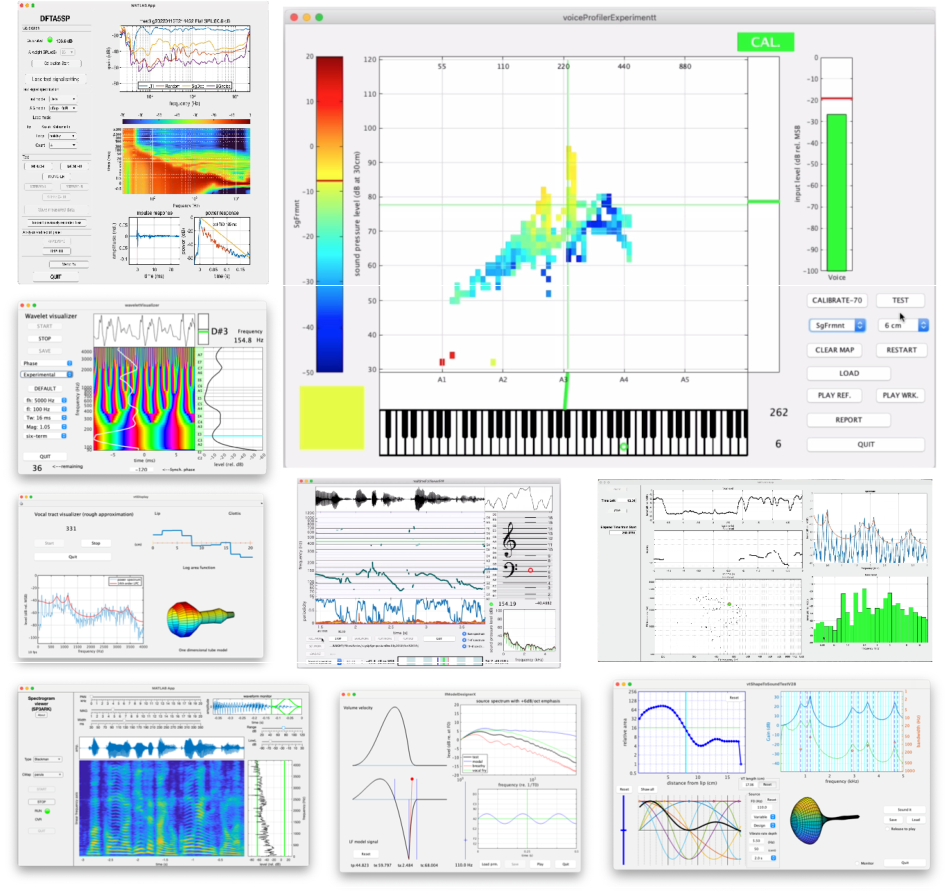}\\ 
\vspace{-2mm}
\caption{Snapshots of application GUIs~\cite{Kawahara2017is,Kawahara2019apsipa,Kawahara2021isST,Kawahara2023smac}. This controller launches applications by selecting the dropdown menu in the application launcher sub-panel.}
\label{fig:toolsFigs}
\end{center}
\end{figure}
Figure~\ref{fig:toolsFigs} shows a collection of GUI snapshots of speech-related applications we developed~\cite{Kawahara2017is,Kawahara2019apsipa,Kawahara2021isST,Kawahara2023smac}.
Selecting an item from the dropdown menu in the application launcher sub-panel launches the application.

The screenshot of Fig.~\ref{fig:acousticCalibTool} shows an interactive and real-time measurement of the sound recording environment using a periodic test signal with a modified pink-noise spectral shape.
A powered loudspeaker (IKmultimedia iLound Micro Monitor) simulates a talker, and a measuring microphone (Earthworks M50) placed 20~cm in front of the loudspeaker acquires the output sound.
The top-right graph shows the smoothed gain (blue line) of the transfer function $H[k]$ and the smoothed disturbing component (red line).
The following graph shows the impulse response waveform (red line) and the smoothed power of the impulse response represented in dB (blue line).

\begin{figure}[tbp]
\begin{center}
\includegraphics[width=\hsize]{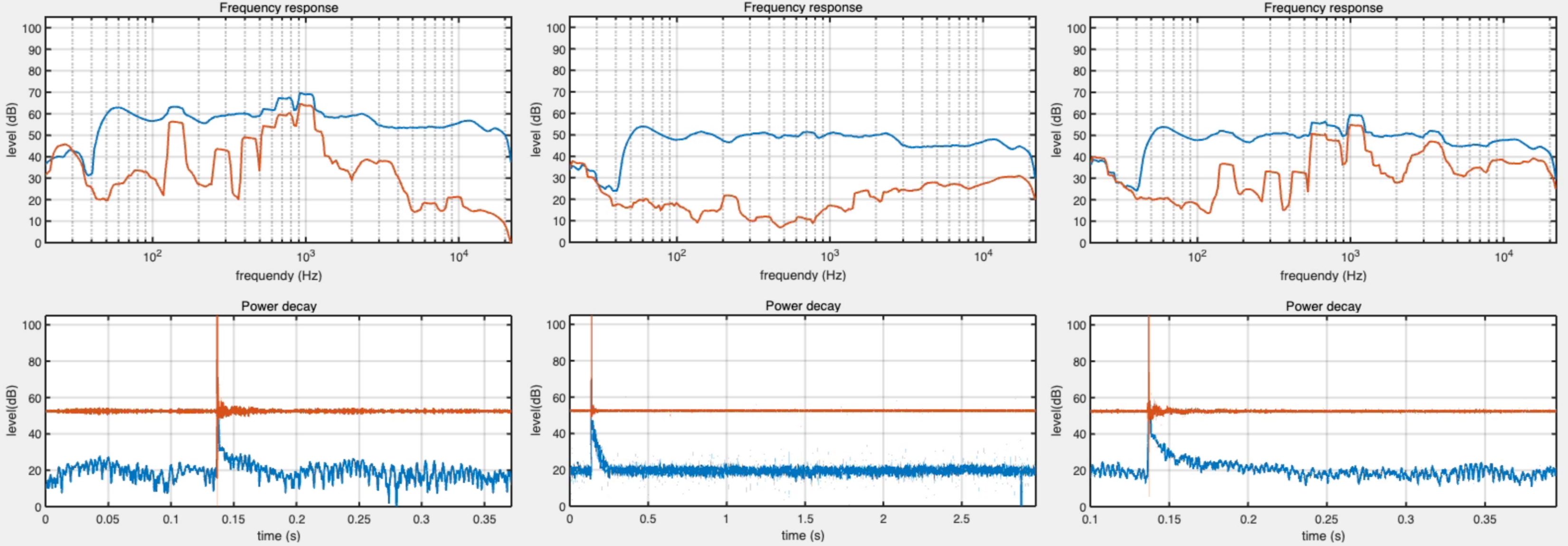}
\caption{Snapshots of monitored acoustic conditions. The left shows when the experimenter disturbed the measurement by one's voice /a/. The center shows the monitored LTI response and background noise using a test sound that is a safeguarded utterance (a Japanese sentence /bakuoN ga ginsekai no kougen ni hirogaru/ ``A roaring sound spreads across the silvery snow-covered plateau.'') spoken by a Japanese male.
These are snapshots excerpted from a demonstration movie.}
\label{fig:acousticMesEx}
\end{center}
\end{figure}
Figure~\ref{fig:acousticMesEx} shows monitored results in different conditions.
These snapshots are excerpts from a demonstration movie\footnote{https://youtu.be/-nxD-8hbCv4}.

The left graph shows when the experimenter disturbs the measurement by producing a loud /a/ voice.
The disturbing component in the red line pushes the LTI estimate at several parts.
The smoothed impulse response power also shows distortions in low-level parts.

The center graph shows the result using a safeguarded read sentence.
The length of the period of the test signal used in this measurement is $2^{17}$.
Figure~\ref{fig:safeGbakuon} in Appendix~\ref{ss:safeguarding} shows the spectra of the original and safeguarded signals.
The measured LTI gain is virtually identical to the result in Fig.~\ref{fig:acousticCalibTool}.
However, the level of disturbing components is higher than Fig.~\ref{fig:acousticCalibTool}.
It is because the power spectral level in the higher frequency region of the safeguarded test signal decreases steeper than that of the pink noise.
Note that the noise floor, due to background noise, masks the smoothed power level of the estimated impulse response other than the initial part of the impulse response.

The right graph shows when the experimenter disturbs the measurement by producing a loud /a/ voice, similar to the left graph.
Similarly, the disturbing component in the red line pushes the LTI estimate at several parts.
There are differences in the shape of the disturbing components in the left and right graphs.
The disturbing level is lower in the low-frequency region and higher in the high-frequency region.
These differences are due to the spectral difference of the test signals.
The pink noise-shaped test signal has higher energy in the high-frequency region than the safeguarded speech sounds.

\section{Other applications}
We developed several tools other than acoustic measurement based on the proposed framework.
This section introduces two applications closely related to each other.

The first one is measurement of voice $f_\mathrm{o}$ (fundamental frequency\footnote{See~\cite{Titze2015asjforum} for the reason why we are using $f_\mathrm{o}$ instead of using F0) response to auditory stimulation~\cite{Kawahara2021isfb,Kawahara2021apsipafb}.}
This research measures the frequency modulation transfer function using the $f_\mathrm{o}$ frequency modulation by the spectrally shaped test signal made from CAPRICEP as the test signal.
Then the $f_\mathrm{o}$ modulation of the produced voice sound as the output. 

The second one is an objective measurement of pitch extractors using the $f_\mathrm{o}$ modulation transfer function~\cite{Kawahara2022isFoMes}.
In this investigation, we replaced the human with the pitch extractor.
The movie that compares 16 pitch extractors using this method is informative\footnote{https://youtu.be/iXnP1tIuVic}.

\section{Discussion and related works}
The proposed framework does not replace existing acoustic measurement methods.
Instead, it provides them additional values and is an efficient computational infrastructure.
For example, exponential swept-sine analyzes nonlinearity in a diagnostic way~\cite{Novak2010eurashipJ}.
Our proposed framework provides information on the target system's behavior handling signals they are designed to use.
In this manner, the proposed framework plays a complementing role in acoustic measurement. 

Reverberation radius~\cite{Mijic2010telforj} represents the distance where the energy of the direct and the reverberant sounds equals.
It is an essential attribute in, for example, classroom acoustics.
Our method enables us to measure the reverberation radius and other acoustic attributes using actual teaching materials in a classroom while teaching students.
Moreover, using the smartphones of each student, simultaneous measurements of each student's listening acoustic conditions are possible.

\section{Conclusions}
We introduced a general framework for measuring acoustic properties such as liner time-invariant (LTI) response, signal-dependent component, and random and time-varying component simultaneously using structured periodic test signals.
The framework also enables music pieces and other sound materials as test signals by ``safeguarding'' them by adding slight deterministic ``noise.''
Measurement using swept-sin, MLS (Maximum Length Sequence), and their variants are special cases of the proposed framework.
We implemented interactive and real-time measuring tools based on this framework and made them open-source.
Furthermore, we applied this framework to assess pitch extractors objectively. 
The proposed framework is general enough and applicable to many other fields.

\section*{Acknowledgment}

KAKENHI by JST 21K19794, 21H03468, 21H00497, 21K11957 supported this line of research.
The authors appreciate for careful and constructive reviewers' comments.

\bibliographystyle{IEEEtran}
\bibliography{kawahara2023APSIPAmultMes}

\appendix


\subsection{Reducing truncation artifacts}\label{ss:truncation}
\begin{figure}[h]
\begin{center}
\includegraphics[width=\hsize]{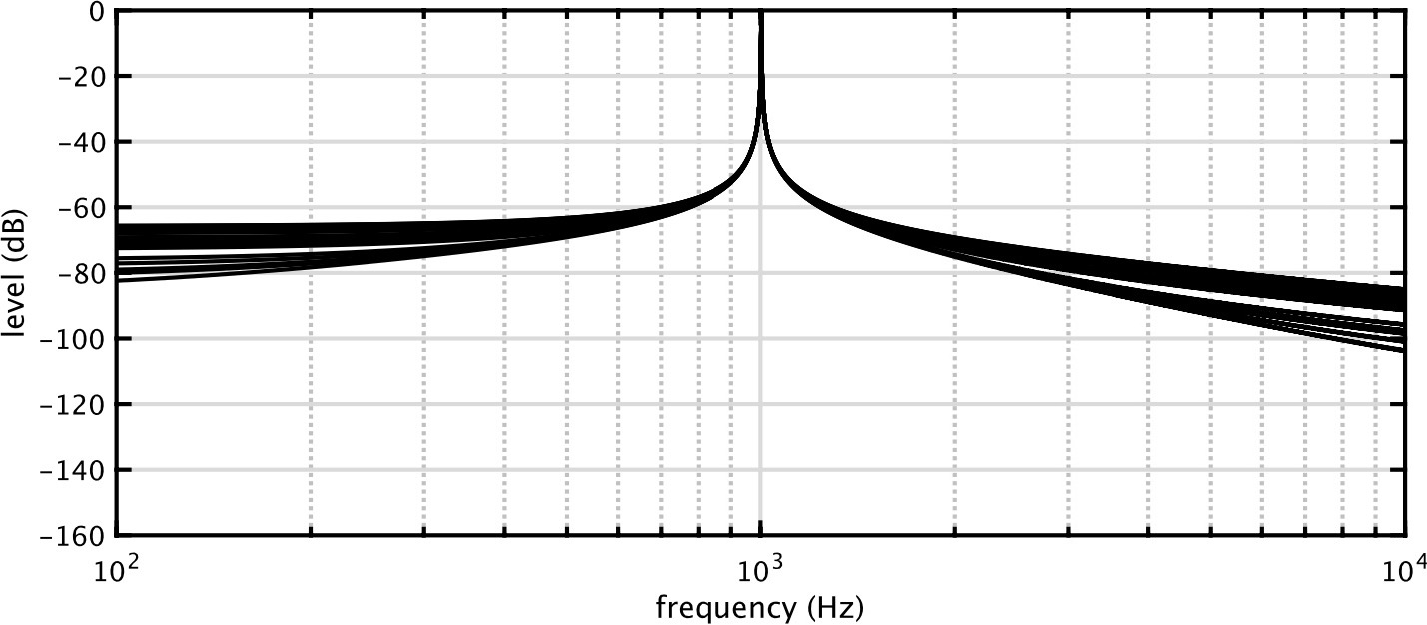}\\ 
\vspace{-2mm}
\caption{Power spectra of truncated signal without any shaping the both ends.\label{effectOfAsyncSigWO}}
\end{center}
\end{figure}
Figure~\ref{effectOfAsyncSigWO} shows an example of signal truncation asynchronous to the signal.
This example uses a sinusoid having a frequency $1000+\pi$~Hz.
As shown in the figure, truncation produces spectral spread.

\begin{figure}[tbp]
\begin{center}
\includegraphics[width=\hsize]{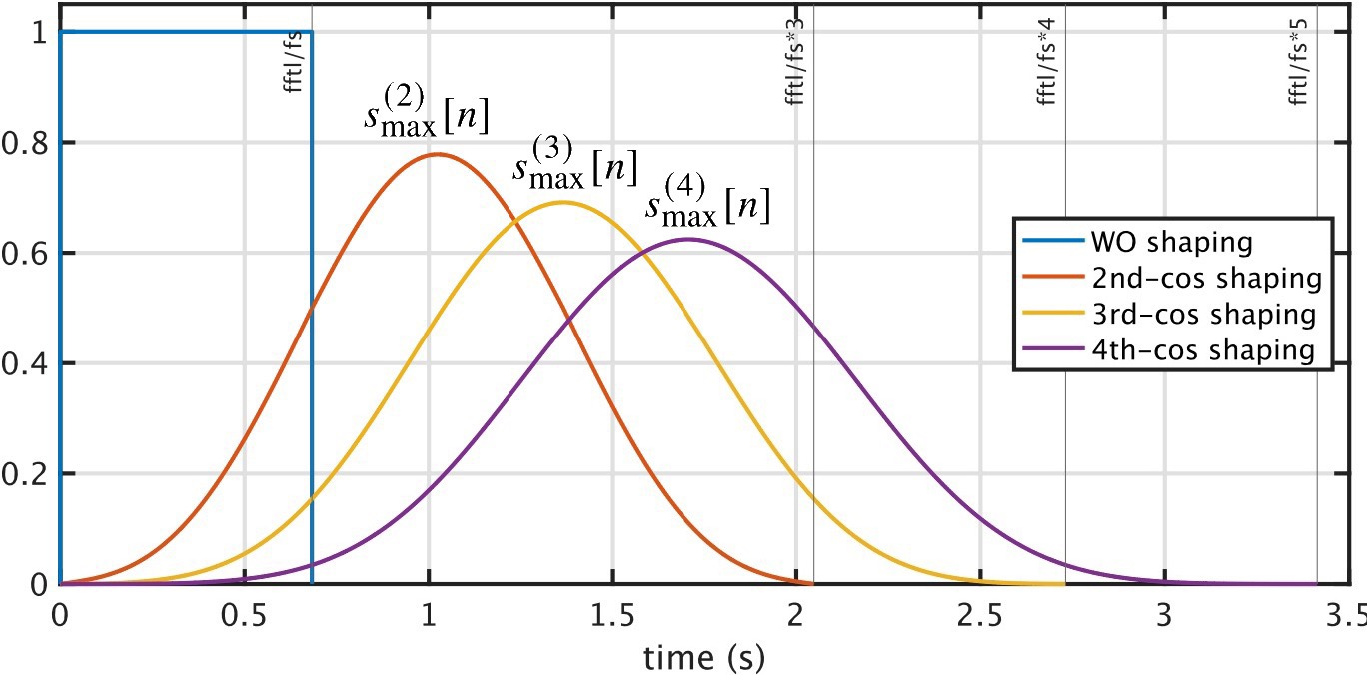}
\caption{Periodicity preservation tests of shaping functions. Plot shows the weighting function shapes. \label{periodCheck}}
\end{center}
\end{figure}
Using more than one cycle of the periodic signal and a weighting function reduces this issue.
The weighting function has to be a constant value one when wrapped.
Figure~\ref{periodCheck} shows such weighting functions.
They are cosine-based window functions convolved with a rectangular function having the period width.
The annotation $s_\mathrm{max}^{(k)}[n]$ indicates that the convolved cosine series has $k$-terms.
The coefficients are numerically optimized to minimize the maximum side-lobe level.

\begin{figure}[tbp]
\begin{center}
\includegraphics[width=\hsize]{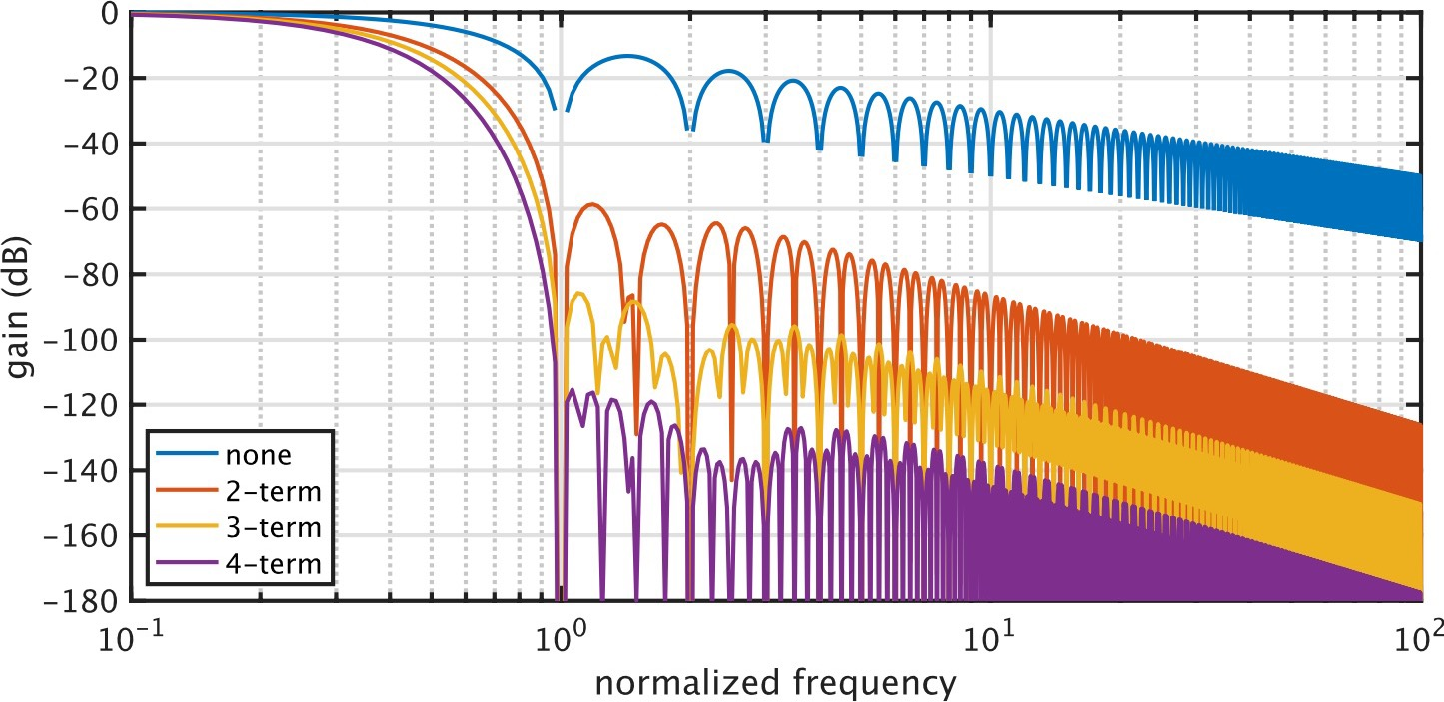}
\caption{Optimized weighting functions made from several cosine series convolved with a rectangular function with the repetition period length.
They are numerically optimezed to minimize the maximum side-lobe level.\label{fig:weightFCosSerMax}}
\end{center}
\end{figure}
Figure~\ref{fig:weightFCosSerMax} shows their frequency responses.

\begin{figure}[tbp]
\begin{center}
\includegraphics[width=\hsize]{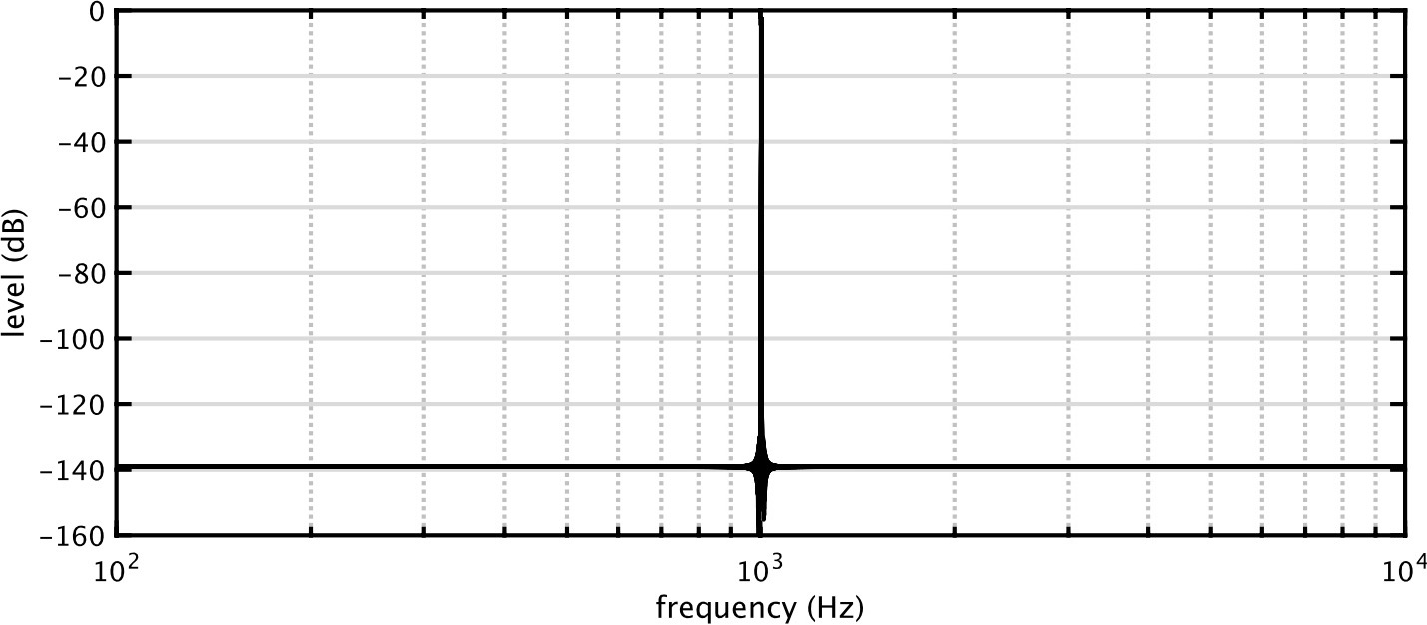}
\caption{Power spectra of truncated signal with optimized shaping weight.\label{effectOfAsyncSigWithOpt}}
\end{center}
\end{figure}
Figure~\ref{effectOfAsyncSigWithOpt} shows the same asynchronous signal truncated using the optimized weighting function $s_\mathrm{max}^{(4)}[n]$.
Note that weighting effectively suppressed the spectral spreads.

\subsection{Phase manipulation function}\label{ss:phaseMan}
\begin{figure}[tbp]
\begin{center}
\includegraphics[width=\hsize]{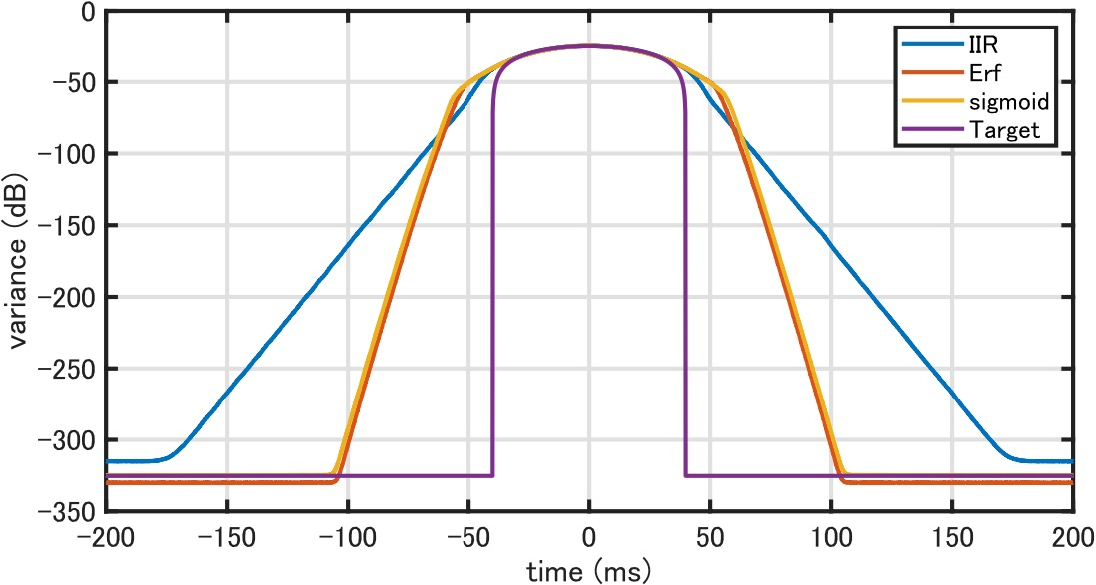}
\caption{Designed outline shapes using a raised cosine shape as the target. The vertical axis represents the variance in logarithmic (dB) scale.\label{fig:shapeDesignResult}}
\end{center}
\end{figure}
Let us start with the result.
Figure~\ref{fig:shapeDesignResult} shows the RMS average of the unit-CAPRICEP designed using a phase manipulation function of the original proposal~\cite{Kawahara2021icassp} (IIR) and the error function, integrated Gaussian function (Erf) and its approximating function sigmoid (sigmoid).

\begin{figure}[tbp]
\begin{center}
\includegraphics[width=\hsize]{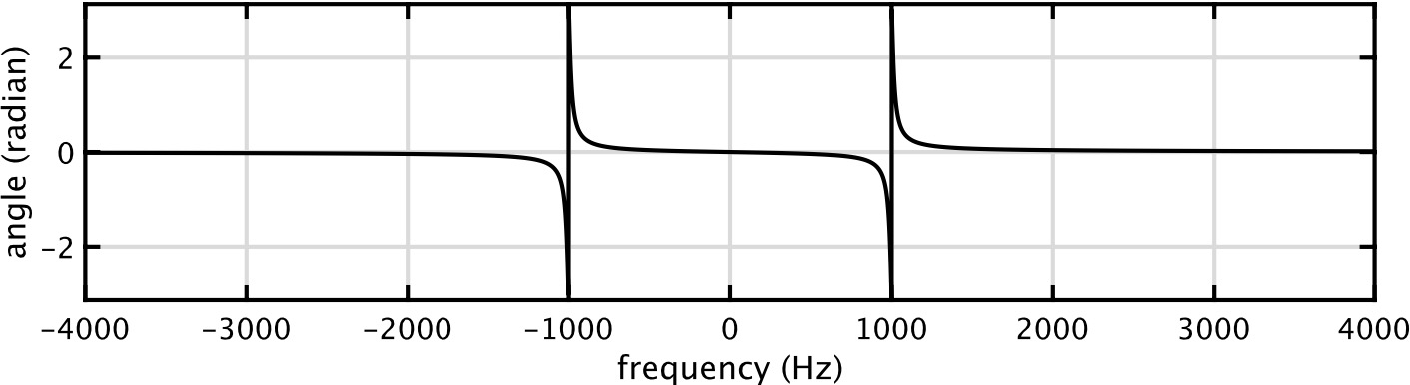}\\
(a)\\
\includegraphics[width=\hsize]{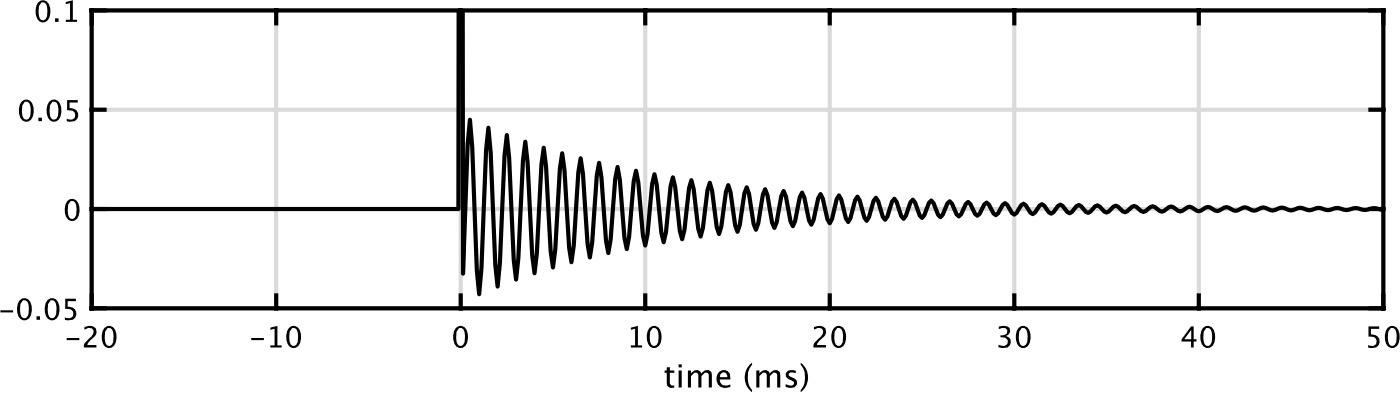}\\
(b)\\
\includegraphics[width=\hsize]{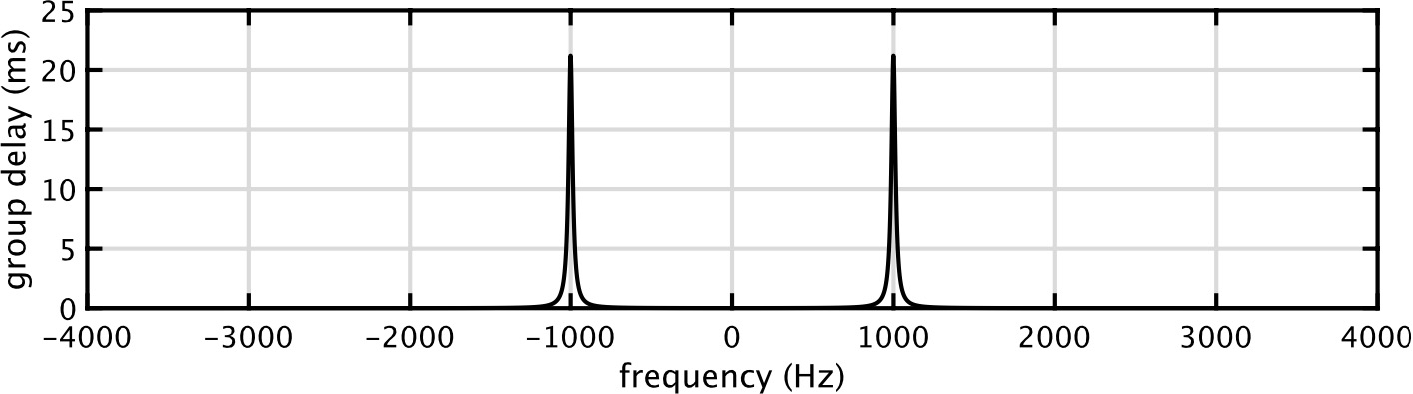}\\
(c)\\
\end{center}
\caption{Attributes of IIR-type all-pass filter. (a) phase-frequency response, (b) impulse response, and (c) group delay response.\label{fig:iirallpass}}
\end{figure}
Figure~\ref{fig:iirallpass} shows an example of relations between phase manipulation, impulse response, and the group delay of the initially proposed phase function~\cite{Kawahara2021icassp}.
It is the phase of an IIR all-pass filter.
This manipulation example and the following examples use phase manipulation at one point, 1000~Hz.
The actual unit-CAPRICEP of 0.25~s effective width consists of about 7000 phase manipulation points.

\begin{figure}[tbp]
\begin{center}
\includegraphics[width=\hsize]{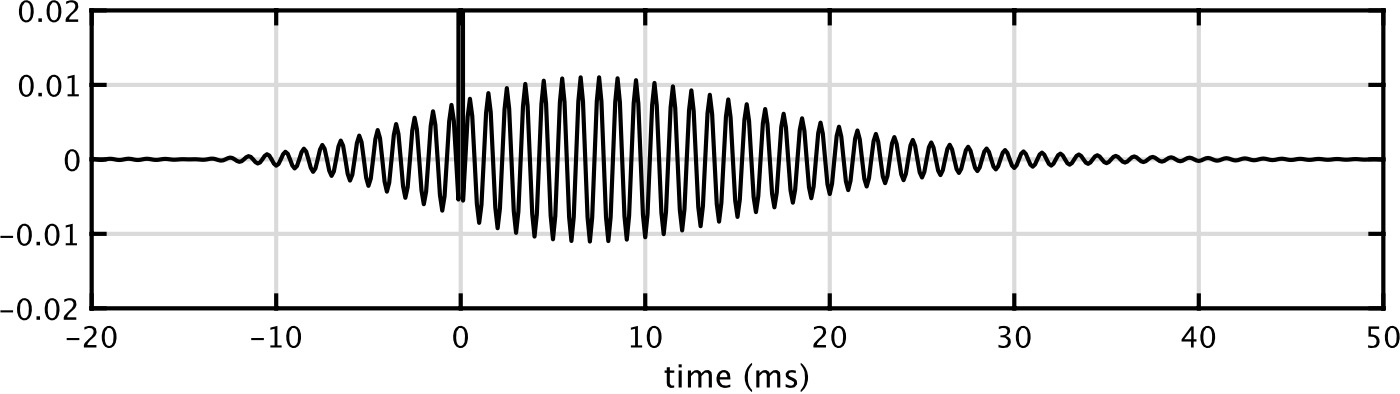}\\
(a)\\
\includegraphics[width=\hsize]{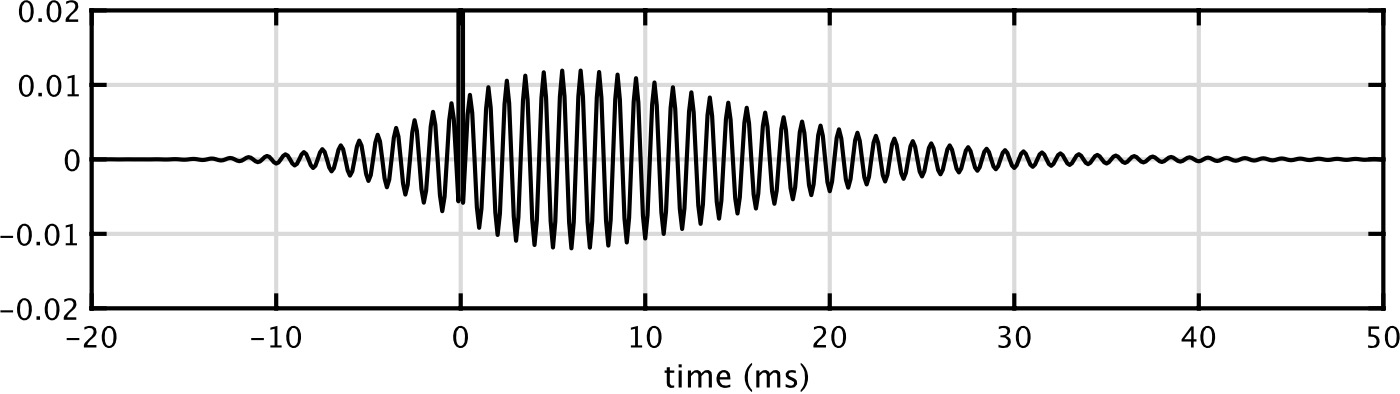}\\
(b)\\
\includegraphics[width=\hsize]{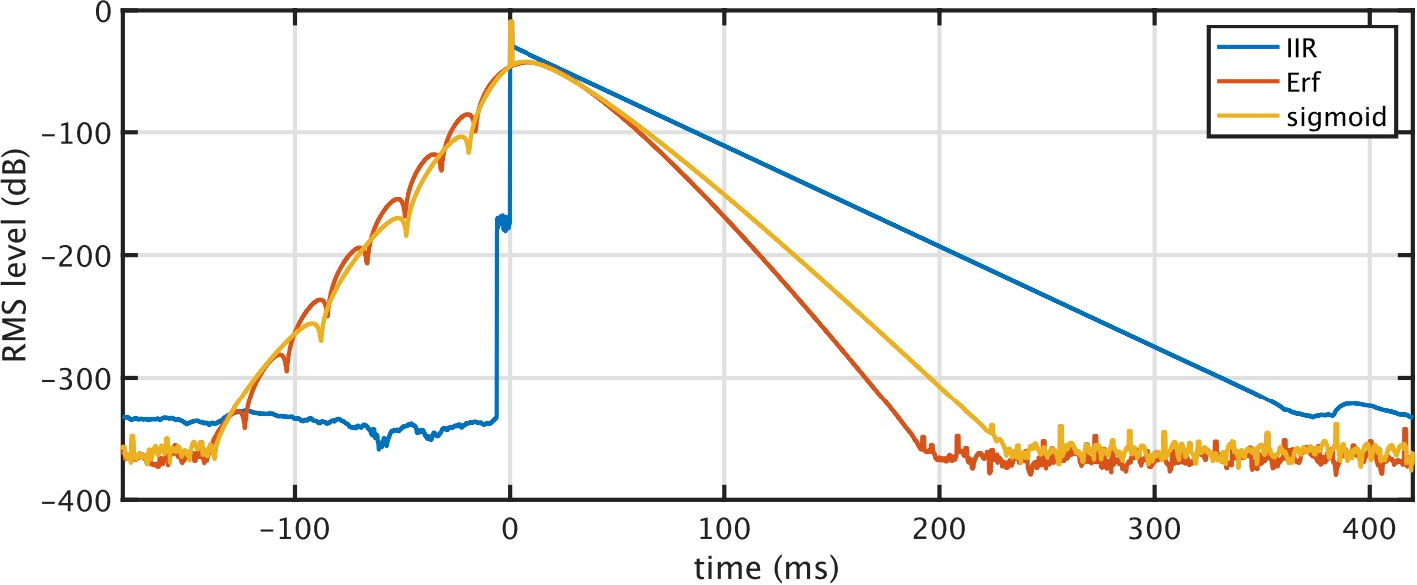}\\
(c)\\
\end{center}
\vspace{-3mm}
\caption{Impulse response of the all-pass filters made from error function (a) and the sigmoid (b). The plot (c) shows smoothed RMS levels (in dB) of the impulse response of the IIR (original CAPRICEP~\cite{Kawahara2021icassp}), the error function (Erf), and the sigmoid (sigmoid). \label{fig:erfallpass}}
\end{figure}
Figure~\ref{fig:erfallpass} shows impulse responses using the error function and the sigmoid.
The bottom plot compares IIR, Erf, and Sigmoid's power decay.
The decay of Erf is the steepest, and we selected Erf to design unit-CAPRICEP in our implementation.

\subsection{Safeguarded signal and its merits}\label{ss:safeguarding}
This is a revised excerpt form~\cite{Kawahara2022ast}.
Let $x[n]$ be a periodic discrete-time signal with a period $L$.
Convolution of $x[n]$ and the impulse response $h[n]$ of the target system yields the output $y[n]$.
Because the signal is periodic, the DFT (Discrete Fourier transform) of $x[n]$ and $y[n]$ segments (their length is $L$) are invariant other than the phase rotation proportional to frequency.
Let $X[k]$ and $Y[k]$ represent their DFT, where $k$, ($k = 0, \ldots, L-1$), is the discrete frequency.
Then, the ratio $Y[k]/X[k]$ is independent of the location of the segment.
This ratio agrees with the DFT $H[k]$ of the impulse response $h[n]$,
where $X[k] \neq 0$ for all $k$ values is the condition of this relation to provide physically meaningful results.

However, this simple solution is sensitive to noise when the absolute value $\left|X[k]\right|$ is very small relative to absolute values $\left|H[k]\right|$ of other $k$ values.
We propose to limit the absolute value $\left|X[k]\right|$ to have larger value than a threshold\footnote{%
In actual implementation we use frequency dependent threshold $\theta_L[k]$ using power spectra of the original signal and the background noise.
We also set the minimum level and low frequency limit for $\theta_L[k]$.}.
We use the following equation to derive the DFT $X_\mathrm{s}[m]$ of the safeguarded signal $\tilde{x}_\mathrm{s}[n]$.
\begin{align}
\!\!X_\mathrm{s}[k] & = \left\{\!\!\begin{array}{lll}
\displaystyle \frac{\theta_L[k]X[k]}{\left|X[k]\right|}  & \mbox{for}  & 0 < \left|X[k]\right| < \theta_L[k]  \\
X[k] & { } & \theta_L \leq \left|X[k]\right| 
\end{array}\right. ,
\end{align}
where we set $X_\mathrm{s}[k] = \theta_L[k]$ when $X[k] = 0$.
Then, we derive the safeguarded transfer function $H_\mathrm{s}[k]$ as follows.
\begin{align}
H_\mathrm{s}[k] & = \frac{Y_\mathrm{s}[k]}{X_\mathrm{s}[k]} ,
\end{align}
where $Y_\mathrm{s}[k]$ represents the DFT of the output of the target system for periodic test signal $\tilde{x}_\mathrm{s}[n]$. 
Because the safeguarded signal $\tilde{x}_\mathrm{s}[n]$ is periodic, we can make \textit{the safeguarded test signal} for acoustic measurement by concatenating it as many times as required.

\begin{figure}[tbp]
\centering
\includegraphics[width=\columnwidth]{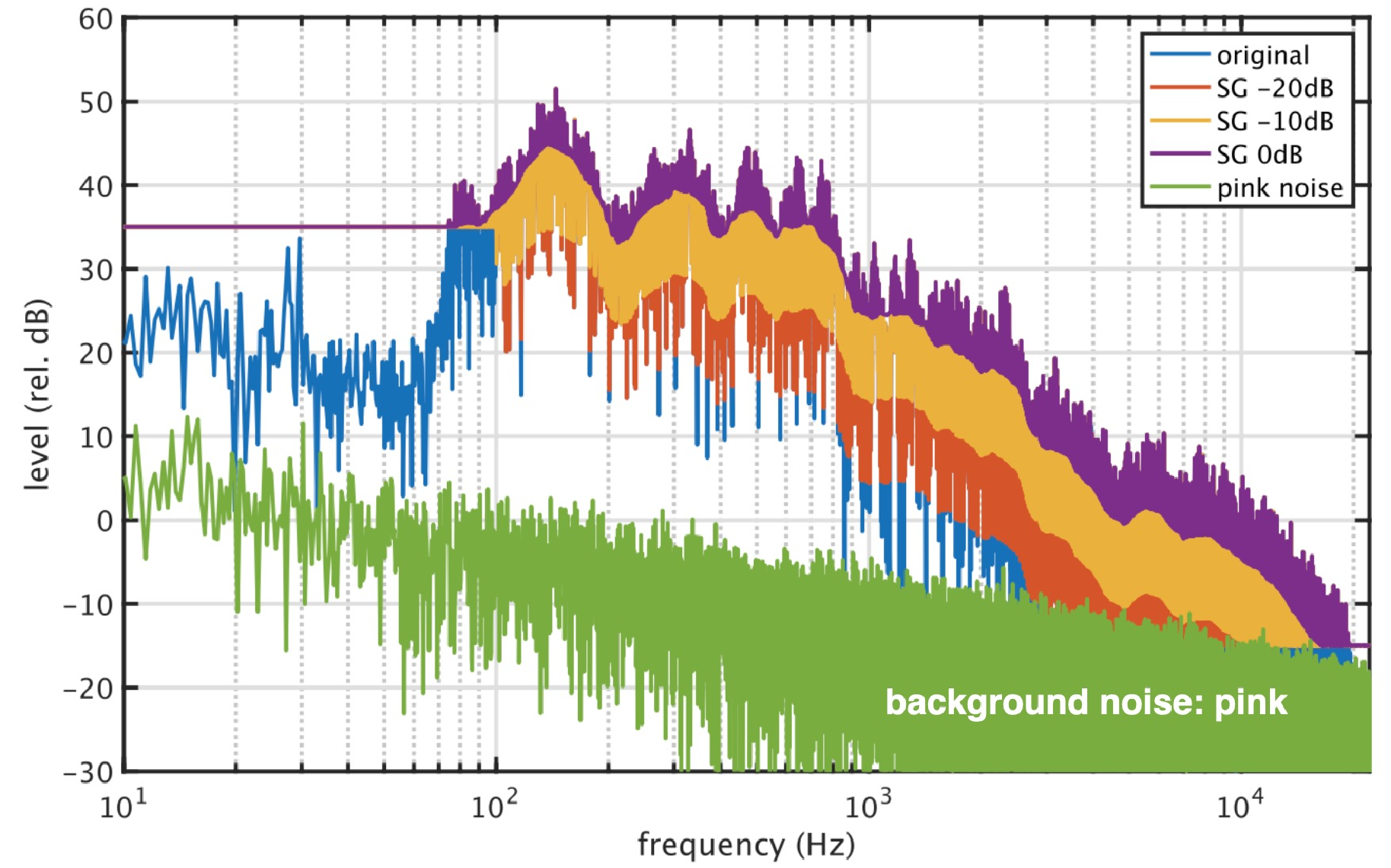}\\ 
\vspace{-2mm}
\caption{Absolute values of the original (a Japanese sentence /bakuon ga ginsekai no kougeN ni hirogaru/ spoken by a male speaker) and safeguarded speech samples.
The signal period is $2^{17}$ samples at 44100~Hz.
Frequency-dependent thresholding uses a smoothed power spectrum with 1/3 octave width as a reference.
The light green line represents the background pink noise for testing noise tolerance.\label{fig:safeGbakuon}}
\end{figure}
\begin{figure}[tbp]
\centering
\includegraphics[width=\columnwidth]{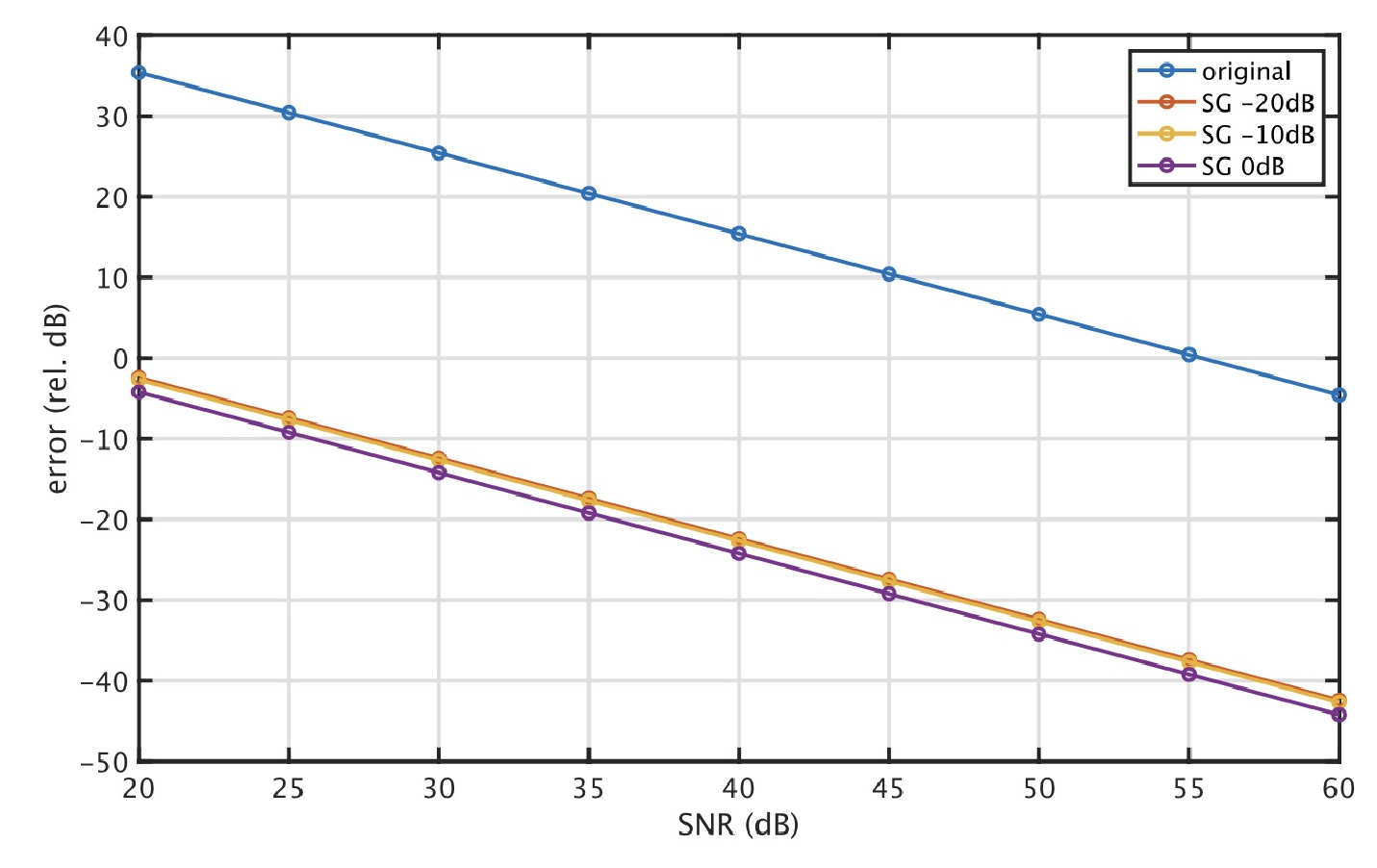}\\
\vspace{-3mm}
\caption{Estimation error of the impulse responses using the original and the safeguarded signals for the test signal.
For noise tolerance test, we added a background pink noise as shown in Fig.~\ref{fig:safeGbakuon}.\label{fig:estimationErr}}
\end{figure}
Figure~\ref{fig:safeGbakuon} shows example of safeguarding.
The level represents the absolute values of discrete Fourier transform of a whole sentence length samples (the original and safeguarded ones).
Figure~\ref{fig:estimationErr} illustrates the merit of safeguarding.

\end{document}